\documentclass[aps,reprint,floatfix,11pt,tightenlines,superscriptaddress,amsmath,amssymb,prl]{revtex4-1}
\usepackage{graphicx}% Include figure files
\usepackage{dcolumn}% Align table columns on decimal point
\usepackage{bm}% bold math
\usepackage{amsmath}

\newcommand{\be}{\begin{eqnarray}}
\newcommand{\ee}{\end{eqnarray}}

\begin{document}

\preprint{ver.2}

\title{\boldmath Reply to ``Comment on `Discovery of slow magnetic fluctuations and critical slowing down in the pseudogap phase of YBa$_2$Cu$_3$O$_y$'\,'' }

\author{Jian Zhang}
\author{Z. F. Ding}
\author{C. Tan}
\author{K. Huang}
\affiliation{State Key Laboratory of Surface Physics, Department of Physics, Fudan University, Shanghai 200433, People's Republic of China}
\author{O. O. Bernal}
\affiliation{Department of Physics and Astronomy, California State University, Los Angeles, California 90032, USA}
\author{P.-C. Ho}
\affiliation{Department of Physics, California State University, Fresno, California 93740,USA}
\author{G. D. Morris}
\affiliation{TRIUMF, Vancouver, BC V6T 2A3, Canada}
\author{A. D. Hillier}
\author{P. K. Biswas}
\author{S. P. Cottrell}
\affiliation{ISIS Facility, STFC Rutherford Appleton Laboratory, Harwell Science and Innovation Campus, Chilton, Didcot, Oxon., UK}
\author{H. Xiang}
\affiliation{State Key Lab for Metal Matrix Composites, Key Lab of Artificial Structures $\&$ Quantum Control (Ministry of Education), Dept.\ of Physics and Astronomy, Shanghai Jiao Tong University, Shanghai 200240, People's Republic of China}
\author{X. Yao}
\affiliation{State Key Lab for Metal Matrix Composites, Key Lab of Artificial Structures $\&$ Quantum Control (Ministry of Education), Dept.\ of Physics and Astronomy, Shanghai Jiao Tong University, Shanghai 200240, People's Republic of China}
\affiliation{Collaborative Innovation Center of Advanced Microstructures, Nanjing 210093, People's Republic of China}
\author{D. E. MacLaughlin}
\email{macl@physics.ucr.edu}
\affiliation{Department of Physics and Astronomy, University of California, Riverside, California 92521, USA}
\author{Lei Shu}
\email{leishu@fudan.edu.cn}
\affiliation{State Key Laboratory of Surface Physics, Department of Physics, Fudan University, Shanghai 200433, People's Republic of China}
\affiliation{Collaborative Innovation Center of Advanced Microstructures, Nanjing 210093, People's Republic of China}
\date{\today}

\begin{abstract}
We reply to the objections raised by J. E. Sonier (arXiv:1706.03023) against our observation of slow magnetic fluctuations and critical slowing down of magnetic fluctuations in the pseudogap phase of YBa$_2$Cu$_3$O$_y$ by zero-field and longitudinal-field muon spin relaxation. 
\end{abstract}

\maketitle

In a recent arXiv post~\cite{Sonier17} (the Comment), J.~E. Sonier has raised objections to our report~\cite{Zhang17} (the Article) of muon spin relaxation ($\mu$SR) experiments that reveal slow magnetic fluctuations and critical slowing down in the pseudogap phase of YBa$_2$Cu$_3$O$_y$ (YBCO). The Comment claims that ``\dots the relaxation data displayed in this study are misleading due to an improper account of the nuclear dipole contribution and a failure to acknowledge the occurrence of muon diffusion'' in analyzing data from zero-field $\mu$SR (ZF-$\mu$SR) experiments. In this Reply we argue that these and other objections raised in the Comment are unfounded.

\textit{Static local field distribution.} We have fit our ZF-$\mu$SR data using the static Gaussian Kubo-Toyabe (GKT) function that describes muon spin relaxation in a randomly-oriented static Gaussian distribution of local magnetic fields~\cite{Hayano79}, damped by a factor~$\exp(-\lambda_\mathrm{ZF}t)$ [Comment, Eqs.~(2) and (3)]. The fitting parameters are the damping rate~$\lambda_\mathrm{ZF}$, due at least in part to a fluctuating component of the muon local field, and the static relaxation rate~$\Delta_\mathrm{ZF}$, where $\Delta_\mathrm{ZF}/\gamma_\mu$ is the rms width of the local field distribution components ($\gamma_\mu$ is the muon gyromagnetic ratio). (Fits using the dynamic GKT function~\cite{Hayano79} to model the effect of muon diffusion at high temperatures are discussed below.) 

The Comment notes that the static GKT function is not an adequate description of relaxation due to nuclear dipolar fields in real materials. This is certainly the case, but it does not affect our study. The static GKT function is only an approximation, but it and calculated ``exact'' relaxation functions (See, e.g., Ref.~\cite{Huang12}) all share a Gaussian-like form for early times. Our results do not involve the origin or the magnitude of the static muon local field distribution, and the values of $\Delta_\mathrm{ZF}$ obtained from our damped GKT fits are simply empirical characterizations of this distribution. Furthermore, for La$_{1.784}$Sr$_{0.216}$CuO$_4$ (LSCO) a static GKT fit is good out to ${\sim}6~\mu$s~\cite{Huang12}. In YBCO the Gaussian relaxation is a factor of two weaker than in LSCO~\cite{Zhang17}, thus extending the early-time range correspondingly. These considerations justify our use of the static GKT form. We also note that the initial parabolic decay (i.e., initial zero slope) of a Gaussian-like relaxation function cannot explain the observed initial nonzero slope due to the exponential component of the relaxation relaxation. 

\textit{Temperature dependence of $\Delta_\mathrm{ZF}$.} Concerning our observation of peaks in $\lambda_\mathrm{ZF}(T)$ (Article, Fig.~2), the Comment claims that the assumption of a tem\-per\-a\-ture-independent $\Delta_\mathrm{ZF}$ is not valid, as it ignores charge order and a structural change near 60~K\@. For $y = 6.985$ a dip in $\Delta_\mathrm{ZF}(T)$ and a concomitant peak in $\lambda_\mathrm{ZF}(T)$ were observed at $\sim$50~K by Sonier \textit{et al.}~\cite{Sonier02b}, and were attributed to the onset of charge ordering at this temperature. 

\begin{figure} [ht]
\centering
\includegraphics[clip=,width=3.25 in]{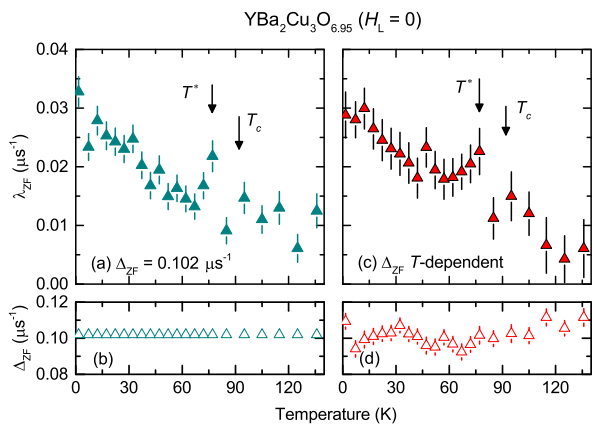} 
\caption{Temperature dependence of zero-field exponential damping rate $\lambda_\mathrm{ZF}$ and static GKT rate $\Delta_\mathrm{ZF}$ for YBa$_2$Cu$_3$O$_{6.95}$. (a) and (b):~Fits with $\Delta_\mathrm{ZF}$ fixed. (c) and (d):~Fits with $\Delta_\mathrm{ZF}$ a free parameter. $T^\ast$ and $T_c$ are the pseudogap and superconducting transition temperatures, respectively.}
\label{6.95}
\end{figure}

We have fit our data from a YBa$_2$Cu$_3$O$_{6.95}$ sample with $\Delta_\mathrm{ZF}$ both fixed and a free parameter, the latter taking into account any temperature dependence. As shown in Fig.~\ref{6.95}, 
both fits yield a peak in $\lambda_\mathrm{ZF}(T)$ at $\sim$80~K near the pseudogap temperature~$T^\ast$, but with no dip or other anomaly in $\Delta_\mathrm{ZF}(T)$ at that temperature with $\Delta_\mathrm{ZF}$ free [Fig.~\ref{6.95}(d)]. (Possible features at $\sim$50~K are much smaller than those reported in Ref.~\cite{Sonier02b}). The absence of a dip is evidence against the charge-ordering scenario for the peak in $\lambda_\mathrm{ZF}(T)$. 

In Fig.~2(b) of the Comment, $\lambda_\mathrm{ZF}(T)$ for $y = 6.985$ from damped static GKT fits with fixed $\Delta_\mathrm{ZF}$ is compared to a peak at $\sim$60~K in the $^{63}$Cu nuclear quadrupole resonance (NQR) linewidth for $y = 7.0$~\cite{GBC00-comm,*GBC00}. This is of course reasonable. But the NQR linewidth peak found for fully-doped YBCO is not observed for slightly lower oxygen content~\cite{KrMe99, KrMe-repl00}. This was noted in Ref.~\cite{Sonier02b} but not in the Comment. Our $y = 6.95$ sample is not fully doped, and its superconducting transition temperature (91~K) and that reported in Ref.~\cite{KrMe-repl00} are about the same. Thus our ZF-$\mu$SR results are consistent with the NQR experiments. Furthermore, Fig.~3 of the Comment shows good agreement between our data for $y = 6.95$ and a peak in $\lambda_\mathrm{ZF}(T)$ for $y = 6.92$ obtained with fixed $\Delta_\mathrm{ZF}$ (fits with $\Delta_\mathrm{ZF}$ free are not shown)~\footnote{Contrary to what is stated in the Comment, $\lambda_\mathrm{ZF}$ data are not shown in Ref.~\cite {Sonier02b} for $y = 0.92$. What are shown for that doping are the rate~$\Lambda$ and stretching power~$p$ from fits of the ``stretched exponential'' relaxation function~$G(t) = \exp[-(\Lambda t)^p]$ that mixes exponential and Gaussian contributions.}. 

\textit{Goodness of fit.} Figure 1 of the Comment shows that assuming a temperature-independent $\Delta_\mathrm{ZF}$ for $y = 6.985$ leads to a late-time discrepancy. Our corresponding fits for $y = 6.95$, shown below in Fig.~\ref{Asy_1}, are much better and do not exhibit such a discrepancy. 
\begin{figure} [ht]
 \begin{center}
 \includegraphics[clip=,width=3.25 in]{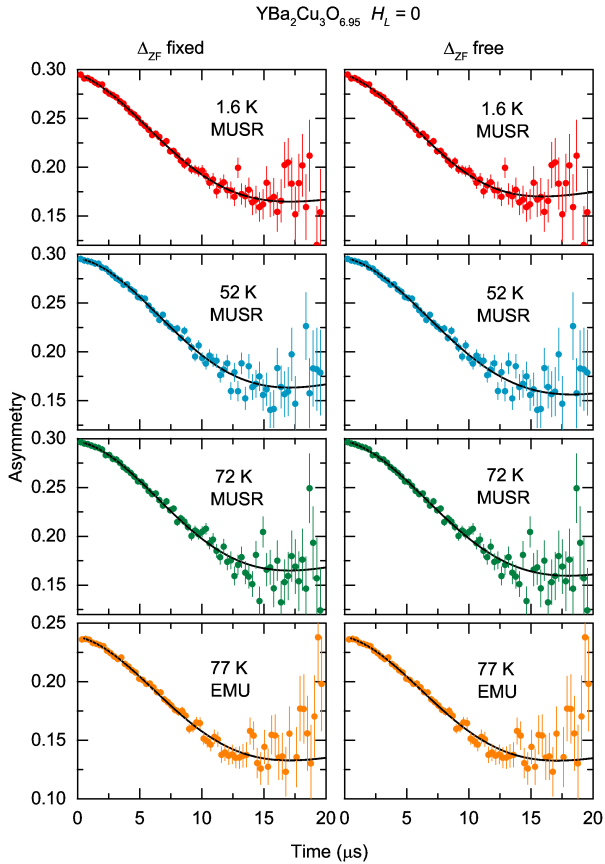}
\caption{Fits of representative ZF-$\mu$SR spectra from YBa$_2$Cu$_3$O$_{6.95}$ for $\Delta_\mathrm{ZF}$ fixed  ($0.1~\mu\text{s}^{-1}$) and free. No late-time discrepancy is observed between data and the fits in either case.}
\label{Asy_1}
 \end{center}
\end{figure}

This behavior, together with that of $\Delta_\mathrm{ZF}(T)$ discussed above and the NQR results~\cite{KrMe99,GBC00}, is evidence that the effect of charge ordering on muon relaxation in the region of the peak is appreciable only in fully-doped YBCO, and is weak or nonexistent (in the region of the peak) in our less-doped $y = 6.95$ sample.

\begin{figure} [ht]
\centering
\includegraphics*[clip=,width=3.25 in]{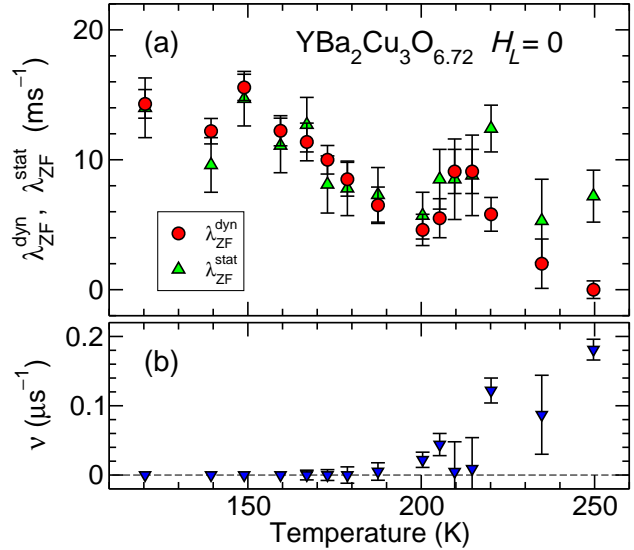} \vspace{-10pt}
\caption{(a)~Comparison of damping rates~$\lambda_\mathrm{ZF}^{\rm dyn}$ and $\lambda_\mathrm{ZF}^{\rm stat}$ from fits of exponentially-damped dynamic and static GKT fit functions, respectively, (see text) to ZF-$\mu$SR data from YBa$_2$Cu$_3$O$_{6.72}$. Peaks are observed for both rates at $\sim$210~K\@. (b)~Muon hopping rate~$\nu(T)$.}
\label{6.72}
\end{figure}

\textit{Muon diffusion.} The Comment claims that muon diffusion above 160~K invalidates the assumption of temperature-independent~$\Delta_\mathrm{ZF}$ for data taken at higher temperatures. We have fit our data from a sample with $y = 6.72$ using the exponentially-damped \textit{dynamic} GKT function~\cite{Hayano79}, which takes muon diffusion into account, for comparison with the static GKT function fits reported in the Article. The corresponding damping rates~$\lambda_\mathrm{ZF}^\mathrm{dyn}(T)$ and $\lambda_\mathrm{ZF}^\mathrm{stat}(T)$ are shown in Fig.~\ref{6.72}(a). 
Both rates exhibit a peak at $\sim$210~K, which is therefore not an artifact of the static GKT fitting function.

The damped dynamic GKT fits are difficult because there are three statistically correlated rate parameters involved: $\Delta_\mathrm{ZF}^\mathrm{dyn}(T)$ (the rms spread of static nuclear dipolar fields between which the muon hops), $\lambda_\mathrm{ZF}^\mathrm{dyn}(T)$, and the muon hopping rate~$\nu(T)$ [Fig.~\ref{6.72}(b)]. Convergence above $\sim$200~K could be achieved only by fixing $\Delta_\mathrm{ZF}^\mathrm{dyn}$. The results for $\nu(T)$ have considerable uncertainty but are similar to those reported in Ref.~\cite{Sonier02b} for $y = 6.67$; in particular, $\nu(T)$ becomes unmeasurably small below $\sim$180~K for both data sets. Thus neither the peak at 160~K in the longitudinal-field damping rate~$\lambda_\mathrm{LF}(T)$ for $y = 6.77$ in 4~mT [Article, Fig.~2(b)] nor the decrease of $\lambda_\mathrm{ZF}(T)$ with increasing temperature between 150~K and 180~K for $y = 6.72$ [Fig.~\ref{6.72}(a) above] can be attributed to muon diffusion. In the Article the onset of low-frequency fluctuations in domains of IUC magnetic order was suggested as the origin of the latter temperature dependence.

As mentioned in the Comment, impurity trapping of diffusing muons can result in a ``peak'' over an intermediate temperature range~\cite{Petzinger80, *BHHL82b}. This is seldom as narrow as our observed peaks, however~\cite{[] [{, Chap.~3.3.}] Schenck85,BHHL82b}. Furthermore, this scenario requires that the rate decrease with increasing temperature below such a peak that be due to significant motional narrowing [$\nu(T) \sim \Delta_\mathrm{ZF}$] of the entire static local field distribution, as the onset of muon diffusion [i.e., the increase of $\nu(T)$] allows muons to find traps. In our case the observed decrease of $\lambda_\mathrm{ZF}(T)$ occurs for temperatures where $\nu(T)$ is negligible [Fig.~\ref{6.72}], and in addition $\Delta_\mathrm{ZF}^\mathrm{dyn}(T)$ is constant in this temperature range (data not shown). The decrease is therefore unlikely to be due muon hopping, and the peak is unlikely to be due to muon trapping.

\textit{Field dependence.} The Comment claims that the field dependence of $\lambda_\mathrm{LF}$ (Article, Fig.~1) deviates from the Redfield functional form [Article, Eq.~(1)] so that the quantitative information is invalid. The statistical uncertainty of the data is large, but the rms local fields and correlation times obtained from fits to the Redfield form have reasonably small errors (Article, Table I). The Article notes that even if this form is not strictly obeyed, the fit parameters are heuristic estimates of the characteristic magnitudes and time scales of the fluctuating local fields, respectively. Absolute parameter values are only used in this sense as order-of-magnitude estimates. Comparison between samples (Article, Table I) shows smooth variation of the parameters with oxygen content. 

\textit{Statistical significance.} The Comment claims that the existence of the peaks (Article, Fig.~2) is questionable. A standard statistical treatment shows, however, that the peaks are significant at the 4--5-sigma level or greater individually and $\sim$8 sigma cumulatively~\footnote{Reference~\cite{Zhang17}, version~2}. This is usually taken as satisfactory.

\textit{Extrinsic effect of field.} The Comment speculates that the observed longitudinal field dependence could be due to changes of the muon beam focus by the field. We note that the effect of increasing the field is to focus the beam, not defocus it, so that this scenario requires beam spots that were off center in the same way for all the samples. (It is routine in $\mu$SR experiments to tune the muon beam channel to center the beam spot on the sample.) Furthermore, in the LAMPF spectrometer at TRIUMF [Article, Figs.~1(a) and (b)] muons that missed the sample were vetoed, whereas in the EMU spectrometer at ISIS [Article, Fig.~1(c)] they stopped in a silver cold finger; nonetheless, results from these very different configurations are similar. Finally, the silver sample used in control LF-$\mu$SR experiments that showed no field dependence (Article, Supplemental Material Fig.~4) was in fact approximately the same size as our YBCO samples. We conclude that an extrinsic origin of the field dependence is highly unlikely.

%\bibliography{reply-v3}

\begin{thebibliography}{14}%
\makeatletter
\providecommand \@ifxundefined [1]{%
 \@ifx{#1\undefined}
}%
\providecommand \@ifnum [1]{%
 \ifnum #1\expandafter \@firstoftwo
 \else \expandafter \@secondoftwo
 \fi
}%
\providecommand \@ifx [1]{%
 \ifx #1\expandafter \@firstoftwo
 \else \expandafter \@secondoftwo
 \fi
}%
\providecommand \natexlab [1]{#1}%
\providecommand \enquote  [1]{``#1''}%
\providecommand \bibnamefont  [1]{#1}%
\providecommand \bibfnamefont [1]{#1}%
\providecommand \citenamefont [1]{#1}%
\providecommand \href@noop [0]{\@secondoftwo}%
\providecommand \href [0]{\begingroup \@sanitize@url \@href}%
\providecommand \@href[1]{\@@startlink{#1}\@@href}%
\providecommand \@@href[1]{\endgroup#1\@@endlink}%
\providecommand \@sanitize@url [0]{\catcode `\\12\catcode `\$12\catcode
  `\&12\catcode `\#12\catcode `\^12\catcode `\_12\catcode `\%12\relax}%
\providecommand \@@startlink[1]{}%
\providecommand \@@endlink[0]{}%
\providecommand \url  [0]{\begingroup\@sanitize@url \@url }%
\providecommand \@url [1]{\endgroup\@href {#1}{\urlprefix }}%
\providecommand \urlprefix  [0]{URL }%
\providecommand \Eprint [0]{\href }%
\providecommand \doibase [0]{http://dx.doi.org/}%
\providecommand \selectlanguage [0]{\@gobble}%
\providecommand \bibinfo  [0]{\@secondoftwo}%
\providecommand \bibfield  [0]{\@secondoftwo}%
\providecommand \translation [1]{[#1]}%
\providecommand \BibitemOpen [0]{}%
\providecommand \bibitemStop [0]{}%
\providecommand \bibitemNoStop [0]{.\EOS\space}%
\providecommand \EOS [0]{\spacefactor3000\relax}%
\providecommand \BibitemShut  [1]{\csname bibitem#1\endcsname}%
\let\auto@bib@innerbib\@empty
%</preamble>
\bibitem [{\citenamefont {{Sonier}}(2017)}]{Sonier17}%
  \BibitemOpen
  \bibfield  {author} {\bibinfo {author} {\bibfnamefont {J.~E.}\ \bibnamefont
  {{Sonier}}},\ }\href@noop {} {\bibfield  {journal} {\bibinfo  {journal}
  {ArXiv e-prints}\ } (\bibinfo {year} {2017})},\ \Eprint
  {http://arxiv.org/abs/1706.03023} {arXiv:1706.\break03023 [cond-mat.supr-con]}
  \BibitemShut {NoStop}%
\bibitem [{\citenamefont {{Zhang}}\ \emph {et~al.}(2017)\citenamefont
  {{Zhang}}, \citenamefont {{Ding}}, \citenamefont {{Tan}}, \citenamefont
  {{Huang}}, \citenamefont {{Bernal}}, \citenamefont {{Ho}}, \citenamefont
  {{Morris}}, \citenamefont {{Hillier}}, \citenamefont {{Biswas}},
  \citenamefont {{Cottrell}}, \citenamefont {{Xiang}}, \citenamefont {{Yao}},
  \citenamefont {{MacLaughlin}},\ and\ \citenamefont {{Shu}}}]{Zhang17}%
  \BibitemOpen
  \bibfield  {author} {\bibinfo {author} {\bibfnamefont {J.}~\bibnamefont
  {{Zhang}}}, \bibinfo {author} {\bibfnamefont {Z.~F.}\ \bibnamefont {{Ding}}},
  \bibinfo {author} {\bibfnamefont {C.}~\bibnamefont {{Tan}}}, \bibinfo
  {author} {\bibfnamefont {K.}~\bibnamefont {{Huang}}}, \bibinfo {author}
  {\bibfnamefont {O.~O.}\ \bibnamefont {{Bernal}}}, \bibinfo {author}
  {\bibfnamefont {P.-C.}\ \bibnamefont {{Ho}}}, \bibinfo {author}
  {\bibfnamefont {G.~D.}\ \bibnamefont {{Morris}}}, \bibinfo {author}
  {\bibfnamefont {A.~D.}\ \bibnamefont {{Hillier}}}, \bibinfo {author}
  {\bibfnamefont {P.~K.}\ \bibnamefont {{Biswas}}}, \bibinfo {author}
  {\bibfnamefont {S.~P.}\ \bibnamefont {{Cottrell}}}, \bibinfo {author}
  {\bibfnamefont {H.}~\bibnamefont {{Xiang}}}, \bibinfo {author} {\bibfnamefont
  {X.}~\bibnamefont {{Yao}}}, \bibinfo {author} {\bibfnamefont {D.~E.}\
  \bibnamefont {{MacLaughlin}}},\ and\ \bibinfo {author} {\bibfnamefont
  {L.}~\bibnamefont {{Shu}}},\ }\href@noop {} {\bibfield  {journal} {\bibinfo
  {journal} {ArXiv e-prints}\ } (\bibinfo {year} {2017})},\ \Eprint
  {http://arxiv.org/abs/1703.06799} {arXiv:1703.06799 [cond-mat.supr-con]}
  \BibitemShut {NoStop}%
\bibitem [{\citenamefont {Hayano}\ \emph {et~al.}(1979)\citenamefont {Hayano},
  \citenamefont {Uemura}, \citenamefont {Imazato}, \citenamefont {Nishida},
  \citenamefont {Yamazaki},\ and\ \citenamefont {Kubo}}]{Hayano79}%
  \BibitemOpen
  \bibfield  {author} {\bibinfo {author} {\bibfnamefont {R.~S.}\ \bibnamefont
  {Hayano}}, \bibinfo {author} {\bibfnamefont {Y.~J.}\ \bibnamefont {Uemura}},
  \bibinfo {author} {\bibfnamefont {J.}~\bibnamefont {Imazato}}, \bibinfo
  {author} {\bibfnamefont {N.}~\bibnamefont {Nishida}}, \bibinfo {author}
  {\bibfnamefont {T.}~\bibnamefont {Yamazaki}},\ and\ \bibinfo {author}
  {\bibfnamefont {R.}~\bibnamefont {Kubo}},\ }\href {\doibase
  10.1103/PhysRevB.20.850} {\bibfield  {journal} {\bibinfo  {journal} {Phys.
  Rev. B}\ }\textbf {\bibinfo {volume} {20}},\ \bibinfo {pages} {850} (\bibinfo
  {year} {1979})}\BibitemShut {NoStop}%
\bibitem [{\citenamefont {Huang}\ \emph {et~al.}(2012)\citenamefont {Huang},
  \citenamefont {Pacradouni}, \citenamefont {Kennett}, \citenamefont {Komiya},\
  and\ \citenamefont {Sonier}}]{Huang12}%
  \BibitemOpen
  \bibfield  {author} {\bibinfo {author} {\bibfnamefont {W.}~\bibnamefont
  {Huang}}, \bibinfo {author} {\bibfnamefont {V.}~\bibnamefont {Pacradouni}},
  \bibinfo {author} {\bibfnamefont {M.~P.}\ \bibnamefont {Kennett}}, \bibinfo
  {author} {\bibfnamefont {S.}~\bibnamefont {Komiya}},\ and\ \bibinfo {author}
  {\bibfnamefont {J.~E.}\ \bibnamefont {Sonier}},\ }\href {\doibase
  10.1103/PhysRevB.85.104527} {\bibfield  {journal} {\bibinfo  {journal} {Phys.
  Rev. B}\ }\textbf {\bibinfo {volume} {85}},\ \bibinfo {pages} {104527}
  (\bibinfo {year} {2012})}\BibitemShut {NoStop}%
\bibitem [{\citenamefont {Sonier}\ \emph {et~al.}(2002)\citenamefont {Sonier},
  \citenamefont {Brewer}, \citenamefont {Kiefl}, \citenamefont {Heffner},
  \citenamefont {Poon}, \citenamefont {Stubbs}, \citenamefont {Morris},
  \citenamefont {Miller}, \citenamefont {Hardy}, \citenamefont {Liang},
  \citenamefont {Bonn}, \citenamefont {Gardner}, \citenamefont {Stronach},\
  and\ \citenamefont {Curro}}]{Sonier02b}%
  \BibitemOpen
  \bibfield  {author} {\bibinfo {author} {\bibfnamefont {J.~E.}\ \bibnamefont
  {Sonier}}, \bibinfo {author} {\bibfnamefont {J.~H.}\ \bibnamefont {Brewer}},
  \bibinfo {author} {\bibfnamefont {R.~F.}\ \bibnamefont {Kiefl}}, \bibinfo
  {author} {\bibfnamefont {R.~H.}\ \bibnamefont {Heffner}}, \bibinfo {author}
  {\bibfnamefont {K.~F.}\ \bibnamefont {Poon}}, \bibinfo {author}
  {\bibfnamefont {S.~L.}\ \bibnamefont {Stubbs}}, \bibinfo {author}
  {\bibfnamefont {G.~D.}\ \bibnamefont {Morris}}, \bibinfo {author}
  {\bibfnamefont {R.~I.}\ \bibnamefont {Miller}}, \bibinfo {author}
  {\bibfnamefont {W.~N.}\ \bibnamefont {Hardy}}, \bibinfo {author}
  {\bibfnamefont {R.}~\bibnamefont {Liang}}, \bibinfo {author} {\bibfnamefont
  {D.~A.}\ \bibnamefont {Bonn}}, \bibinfo {author} {\bibfnamefont {J.~S.}\
  \bibnamefont {Gardner}}, \bibinfo {author} {\bibfnamefont {C.~E.}\
  \bibnamefont {Stronach}},\ and\ \bibinfo {author} {\bibfnamefont {N.~J.}\
  \bibnamefont {Curro}},\ }\href {\doibase 10.1103/PhysRevB.66.134501}
  {\bibfield  {journal} {\bibinfo  {journal} {Phys. Rev. B}\ }\textbf {\bibinfo
  {volume} {66}},\ \bibinfo {pages} {134501} (\bibinfo {year}
  {2002})}\BibitemShut {NoStop}%
\bibitem [{\citenamefont {Gr\'evin}\ \emph
  {et~al.}(2000{\natexlab{a}})\citenamefont {Gr\'evin}, \citenamefont
  {Berthier},\ and\ \citenamefont {Collin}}]{GBC00-comm}%
  \BibitemOpen
  \bibfield  {author} {\bibinfo {author} {\bibfnamefont {B.}~\bibnamefont
  {Gr\'evin}}, \bibinfo {author} {\bibfnamefont {Y.}~\bibnamefont {Berthier}},\ and\ \bibinfo {author} {\bibfnamefont {G.}~\bibnamefont {Collin}},\ }\href
  {\doibase 10.1103/PhysRevLett.84.1636} {\bibfield  {journal} {\bibinfo
  {journal} {Phys. Rev. Lett.}\ }\textbf {\bibinfo {volume} {84}},\ \bibinfo
  {pages} {1636} (\bibinfo {year} {2000}{\natexlab{a}})}\BibitemShut {NoStop}%
\bibitem [{\citenamefont {Gr\'evin}\ \emph
  {et~al.}(2000{\natexlab{b}})\citenamefont {Gr\'evin}, \citenamefont
  {Berthier},\ and\ \citenamefont {Collin}}]{GBC00}%
  \BibitemOpen
  \bibfield  {author} {\bibinfo {author} {\bibfnamefont {B.}~\bibnamefont
  {Gr\'evin}}, \bibinfo {author} {\bibfnamefont {Y.}~\bibnamefont {Berthier}},\ and\ \bibinfo {author} {\bibfnamefont {G.}~\bibnamefont {Collin}},\ }\href
  {\doibase 10.1103/PhysRevLett.85.1310} {\bibfield  {journal} {\bibinfo
  {journal} {Phys. Rev. Lett.}\ }\textbf {\bibinfo {volume} {85}},\ \bibinfo
  {pages} {1310} (\bibinfo {year} {2000}{\natexlab{b}})}\BibitemShut {NoStop}%
\bibitem [{\citenamefont {Kr\"amer}\ and\ \citenamefont
  {Mehring}(1999)}]{KrMe99}%
  \BibitemOpen
  \bibfield  {author} {\bibinfo {author} {\bibfnamefont {S.}~\bibnamefont
  {Kr\"amer}}\ and\ \bibinfo {author} {\bibfnamefont {M.}~\bibnamefont
  {Mehring}},\ }\href {\doibase 10.1103/PhysRevLett.83.396} {\bibfield
  {journal} {\bibinfo  {journal} {Phys. Rev. Lett.}\ }\textbf {\bibinfo
  {volume} {83}},\ \bibinfo {pages} {396} (\bibinfo {year} {1999})}\BibitemShut
  {NoStop}%
\bibitem [{\citenamefont {Kr\"amer}\ and\ \citenamefont
  {Mehring}(2000)}]{KrMe-repl00}%
  \BibitemOpen
  \bibfield  {author} {\bibinfo {author} {\bibfnamefont {S.}~\bibnamefont
  {Kr\"amer}}\ and\ \bibinfo {author} {\bibfnamefont {M.}~\bibnamefont
  {Mehring}},\ }\href {\doibase 10.1103/PhysRevLett.84.1637} {\bibfield
  {journal} {\bibinfo  {journal} {Phys. Rev. Lett.}\ }\textbf {\bibinfo
  {volume} {84}},\ \bibinfo {pages} {1637} (\bibinfo {year}
  {2000})}\BibitemShut {NoStop}%
\bibitem [{Note1()}]{Note1}%
  \BibitemOpen
  \bibinfo {note} {Contrary to what is stated in the Comment, 
  $\lambda_\mathrm{ZF}$ data are not shown in Ref.~\cite {Sonier02b} for 
  $y = 0.92$. What are shown for that doping are the rate~$\Lambda $ and
  stretching power~$p$ from fits of the ``stretched exponential'' relaxation function~$G(t) = \protect \qopname \relax
  o{exp}[-(\Lambda t)^p]$ that mixes exponential and Gaussian
  contributions.}\BibitemShut {Stop}%
\bibitem [{\citenamefont {Petzinger}(1980)}]{Petzinger80}%
  \BibitemOpen
  \bibfield  {author} {\bibinfo {author} {\bibfnamefont {K.}~\bibnamefont
  {Petzinger}},\ }\href {\doibase
  http://dx.doi.org/10.1016/0375-9601(80)90121-8} {\bibfield  {journal}
  {\bibinfo  {journal} {Phys. Lett. A}\ }\textbf {\bibinfo {volume} {75}},\
  \bibinfo {pages} {225 } (\bibinfo {year} {1980})}\BibitemShut {NoStop}%
\bibitem [{\citenamefont {Boekema}\ \emph {et~al.}(1982)\citenamefont
  {Boekema}, \citenamefont {Heffner}, \citenamefont {Hutson}, \citenamefont
  {Leon}, \citenamefont {Schillaci}, \citenamefont {Kossler}, \citenamefont
  {Numan},\ and\ \citenamefont {Dodds}}]{BHHL82b}%
  \BibitemOpen
  \bibfield  {author} {\bibinfo {author} {\bibfnamefont {C.}~\bibnamefont
  {Boekema}}, \bibinfo {author} {\bibfnamefont {R.~H.}\ \bibnamefont
  {Heffner}}, \bibinfo {author} {\bibfnamefont {R.~L.}\ \bibnamefont {Hutson}},
  \bibinfo {author} {\bibfnamefont {M.}~\bibnamefont {Leon}}, \bibinfo {author}
  {\bibfnamefont {M.~E.}\ \bibnamefont {Schillaci}}, \bibinfo {author}
  {\bibfnamefont {W.~J.}\ \bibnamefont {Kossler}}, \bibinfo {author}
  {\bibfnamefont {M.}~\bibnamefont {Numan}},\ and\ \bibinfo {author}
  {\bibfnamefont {S.~A.}\ \bibnamefont {Dodds}},\ }\href {\doibase
  10.1103/PhysRevB.26.2341} {\bibfield  {journal} {\bibinfo  {journal} {Phys.
  Rev. B}\ }\textbf {\bibinfo {volume} {26}},\ \bibinfo {pages} {2341}
  (\bibinfo {year} {1982})}\BibitemShut {NoStop}%
\bibitem [{\citenamefont {Schenck}(1985)}]{Schenck85}%
  \BibitemOpen
  \bibfield  {author} {\bibinfo {author} {\bibfnamefont {A.}~\bibnamefont
  {Schenck}},\ }\href@noop {} {\emph {\bibinfo {title} {Muon Spin Rotation
  Spectroscopy}}},\ edited by\ \bibinfo {editor} {\bibfnamefont {E.~W.~J.}\
  \bibnamefont {Mitchell}}\ (\bibinfo  {publisher} {Adam Hilger Ltd},\ \bibinfo
  {address} {Bristol},\ \bibinfo {year} {1985})\BibitemShut {NoStop}%
\bibitem [{Note2()}]{Note2}%
  \BibitemOpen
  \bibinfo {note} {Reference~\cite {Zhang17}, version~2}\BibitemShut {NoStop}%
\end{thebibliography}

\end{document}